\documentclass[twocolumn,aps,prl,tightenlines,amsmath,amssymb,superscriptaddress]{revtex4} 
\usepackage{hyperref} 
\usepackage{graphicx} 
\usepackage{times} 
\usepackage{amssymb} 
\usepackage{amsfonts} 
\newcommand{\dg}{{\dagger}} 
\newcommand{\pdg}{{\vphantom\dagger}} 
\newcommand{\px}{{p_x}} 
\newcommand{\bQ}{{\bf Q}} 
\newcommand{\bp}{{\bf p}} 
\newcommand{\py}{{p_y}} 
\newcommand{\bea}{ 
\begin{eqnarray}
	} 
	\newcommand{\eea}{ 
\end{eqnarray}
} 
\newcommand{\bG}{{\bf G}} 
\newcommand{\bK}{{\bf K}}

\begin{document} 
\title{Band structures of bilayer graphene superlattices} 
\author{ \footnote{These authors contributed equally.}Matthew Killi} \affiliation{Department of Physics, University of Toronto, Toronto, Ontario, Canada M5S 1A7} 
\author{$^*$Si Wu} 
\affiliation{Department of Physics, University of Toronto, Toronto, Ontario, Canada M5S 1A7} 
\author{Arun Paramekanti}
\affiliation{Department of Physics, University of Toronto, Toronto, Ontario, Canada M5S 1A7} \affiliation{Canadian Institute for Advanced Research, Toronto, Ontario, M5G 1Z8, Canada} \affiliation{Department of Physics, Indian Institute of Science, Bangalore, India 560 012} 
\begin{abstract}
	We formulate a low energy effective Hamiltonian to study superlattices in bilayer graphene (BLG) using a minimal model which supports quadratic band touching points. We show that a one dimensional (1D) periodic modulation of the chemical potential or the electric field perpendicular to the layers leads to the generation of zero-energy anisotropic massless Dirac fermions and finite energy Dirac points with tunable velocities. The electric field superlattice maps onto a coupled chain model comprised of 'topological' edge modes. 2D superlattice modulations are shown to lead to gaps on the mini-Brillouin zone boundary but do not, for certain symmetries, gap out the quadratic band touching point. Such potential variations, induced by impurities and rippling in biased BLG, could lead to subgap modes which are argued to be relevant to understanding transport measurements. 
\end{abstract}

\maketitle

Superlattices provide a route to band structure engineering in semiconductors \cite{Tsu}. In graphene \cite{NetoRMP}, a superlattice (SL) potential has been shown to lead to anisotropic Fermi velocity renormalization \cite{Park1}, and generation of new Dirac points in the spectrum \cite{Park2,Brey,Barbier0,Barbier1} resulting from the chiral nature of massless Dirac excitations. Such graphene SLs have been studied by epitaxial growth of graphene on Ir(111) surface \cite{Plet,Rusponi}. Superlattice effects have also been studied in a topological insulator in proximity to a helical spin density wave \cite{DasSarma}, and in graphene subject to a magnetic SL \cite{Dell'Anna,Masir}. However, apart from transfer matrix studies of 1D Kr\"onig-Penney models \cite{Barbier1,Barbier2}, SLs in bilayer graphene (BLG) have not been carefully explored.

Besides band structure engineering, there is a second motivation to study such BLG SLs. On theoretical grounds, BLG is an attractive candidate for transistor applications since it has a tunable gap which varies in proportion to the electric field perpendicular to the layers \cite{McCann1,McCann2}. However, transport measurements on BLG samples do not show the strong suppression of conductance at low temperatures  expected on theoretical grounds \cite{McCann1,McCann2} or from optical absorption measurements  \cite{ZhangY}.  Instead, the transport data shows evidence for variable range hopping conduction \cite{Oostinga,Tay,Miyazaki} or a suppressed band gap \cite{Tay, Xia}. It has been proposed that the observed excess conductance arises from edge states \cite{LiJian}, but transport measurements in a Corbino geometry do not support this scenario \cite{Yan}, suggesting the existence of disorder-induced low energy modes in the \textit{bulk}. To the extent that disorder potentials can be decomposed into Fourier components, we expect to learn something useful about disordered BLG by studying the simpler problem of periodic potential modulations in BLG.

In this Letter, we study the band structures of BLG SLs, arising from periodic modulations of the chemical potential and the bias, using an effective low energy Hamiltonian. Our main results are the following. (i) Although the minimal model of BLG has {\it quadratic} band touching points, we find, remarkably, that a weak 1D chemical potential modulation leads to the generation of {\it linearly} dispersing massless Dirac fermions with a tunable and anisotropic velocity. These Dirac fermion excitations are robust and rely on the chiral nature of the BLG quasiparticles. Beyond a critical modulation amplitude, these Dirac modes get gapped out. (ii) An electric field SL is shown to support linearly dispersing massless Dirac fermions and finite energy Dirac points which survive even for strong modulations. We provide a picture for these modes within a novel coupled chain model of `topological' edge states. (iii) For 2D SLs, we show that for chemical potential and electric field SLs the quadratic band touching points are protected for symmetric SLs with $C_4$ or $C_6$ symmetry. (iv) We compute the density of states for biased BLG with superimposed 1D potential modulations, and find a plethora of subgap modes which we argue are important for understanding transport data. While our results on 1D SLs overlap with work on Kr\"onig-Penney models \cite{Barbier2,Barbier3}, our analysis provides simpler insights, highlights the role of the quasiparticle chirality, and is applied here to more general potential profiles as well as to 2D SLs. 

{\it Effective Hamiltonian approach. ---} The low energy Hamiltonian for Bernal-stacked BLG can be obtained by expanding its minimal tight binding spectrum near one of the Brillouin zone corners ($\bK$ points) \cite{McCann1}. When the bias (i.e., interlayer potential difference) is not too large, $|\Delta|\ll t_{\perp}$, we find ${\mathcal H}=\psi^{\dagger}\hat{H}\psi$ \cite{McCann1}, where 
\begin{equation}
	\label{Hred} \hat{H}=-\frac{v_F^2}{t_{\perp}}\left( 
	\begin{array}{cc}
		0 & (\pi^\dagger)^2 \\
		\pi^2 & 0 
	\end{array}
	\right)+\left( 
	\begin{array}{cc}
		V_1({\bf x}) & 0 \\
		0 & V_2({\bf x}) 
	\end{array}
	\right), 
\end{equation}
and $\psi^{T}=(a_{\bf x},b_{\bf x})$, with $a$ ($b$) being the electron operator on the top (bottom) layer. Here, $\pi \! =\! -i 
\partial_x \! + \! 
\partial_y$, $v_F \! =\! \sqrt{3}td/2 \! \approx \! 10^6$~m/s is the Fermi velocity, $t \! \approx \! 3$~eV is the nearest neighbor hopping integral, $d \! \approx \! 2.46$ \AA \, is the distance between neighboring atoms on the same sublattice, $V_{1,2}$ are the potentials on each layer, and $t_{\perp} \!\approx \! 0.15 t$ is the interlayer coupling. Unless stated, we set $t\!\!=\!\!d\!\!=\!\!1$. We will ignore inter-valley scattering assuming the potentials are varying slowly on the scale of $d$, so that identical physics is expected around the other valley (at $-\bK$). Such an approach has been successfully used to study SLs in monolayer graphene \cite{Park1,Park2}.

To diagonalize $H_{\rm kin}$, we Fourier transform and then make a unitary transformation $a_{\bf p} \!=\! (\alpha_{\bf p} \!+\! \beta_{\bf p})/\sqrt{2}$, $b_{\bf p} \!=\! {\rm e}^{2 i \theta_{\bf p}} (\alpha_{\bf p} \!-\! \beta_{\bf p})/\sqrt{2}$, where $\cos\theta_{{\bf p}}\!=\!p_x/p$ and $p\!=\!\sqrt{p_x^2+p_y^2}$. This leads to $H_{\rm kin}\!=\!\sum_{\bf p}\left(\varepsilon_{e}({\bf p})\beta^{\dagger}_{\bf p} \beta^\pdg_{\bf p} \!+\! \varepsilon_{h}({\bf p})\alpha^{\dagger}_{\bf p}\alpha^\pdg_{\bf p}\right)$. Here $\varepsilon_{e,h}({\bf p})\!=\!\pm p^2/2 m^*$ are energies of electron (hole) states, with an effective mass $m^* \!\equiv\! t_{\perp}/(2 v_F^2)$. This minimal model supports quadratic band touching points at $\pm \bK$.

When $V_{1,2}({\bf x})$ are periodic, we can also Fourier transform the SL potential to obtain $H_{\rm SL}=\sum_{{\bf p},{\bf G}}\Psi^{\dagger}({\bf p}) W_{{\bf p},{\bf G}} \Psi({\bf p}-{\bf G})$, where 
\begin{eqnarray}
	\label{sl} W_{{\bf p},{\bf G}}\!\!=\!\!\frac{1}{2} \left( 
	\begin{array}{cc}
		\! V_1({\bf G})\!+\!V_2({\bf G}){\rm e}^{2i\theta} & V_1({\bf G})\!-\!V_2({\bf G}){\rm e}^{2i\theta} \! \\
		\! V_1({\bf G})\!-\!V_2({\bf G}){\rm e}^{2i\theta} & V_1({\bf G})\!+\!V_2({\bf G}){\rm e}^{2i\theta} \! 
	\end{array}
	\right)\!\!, \label{W} 
\end{eqnarray}
$\Psi^\dg({\bf p})\!=\!(\alpha^\dg_{\bf p},\beta^\dg_{\bf p})$, and $\theta \!\equiv\! \theta_{{\bf p}-{\bf G}}\!-\!\theta_{{\bf p}}$ is the angle between momenta ${\bf p}\!-\!{\bf G}$ and ${\bf p}$. Our aim is to understand the band structures of SLs described by $H_{\rm kin}+H_{\rm SL}$. We will study 1D SLs with period $\lambda$ along $\hat{y}$, so that the reciprocal lattice vectors, $\{\bG\}$, are integer multiples of $\bQ=(0,2\pi/\lambda)$, and the mini Brillouin zone (MBZ) boundaries are at $p_y=\pm \pi/\lambda$. We will also study 2D SLs.

\begin{figure}
	[b] \centering 
	\includegraphics[width=.23 
	\textwidth]{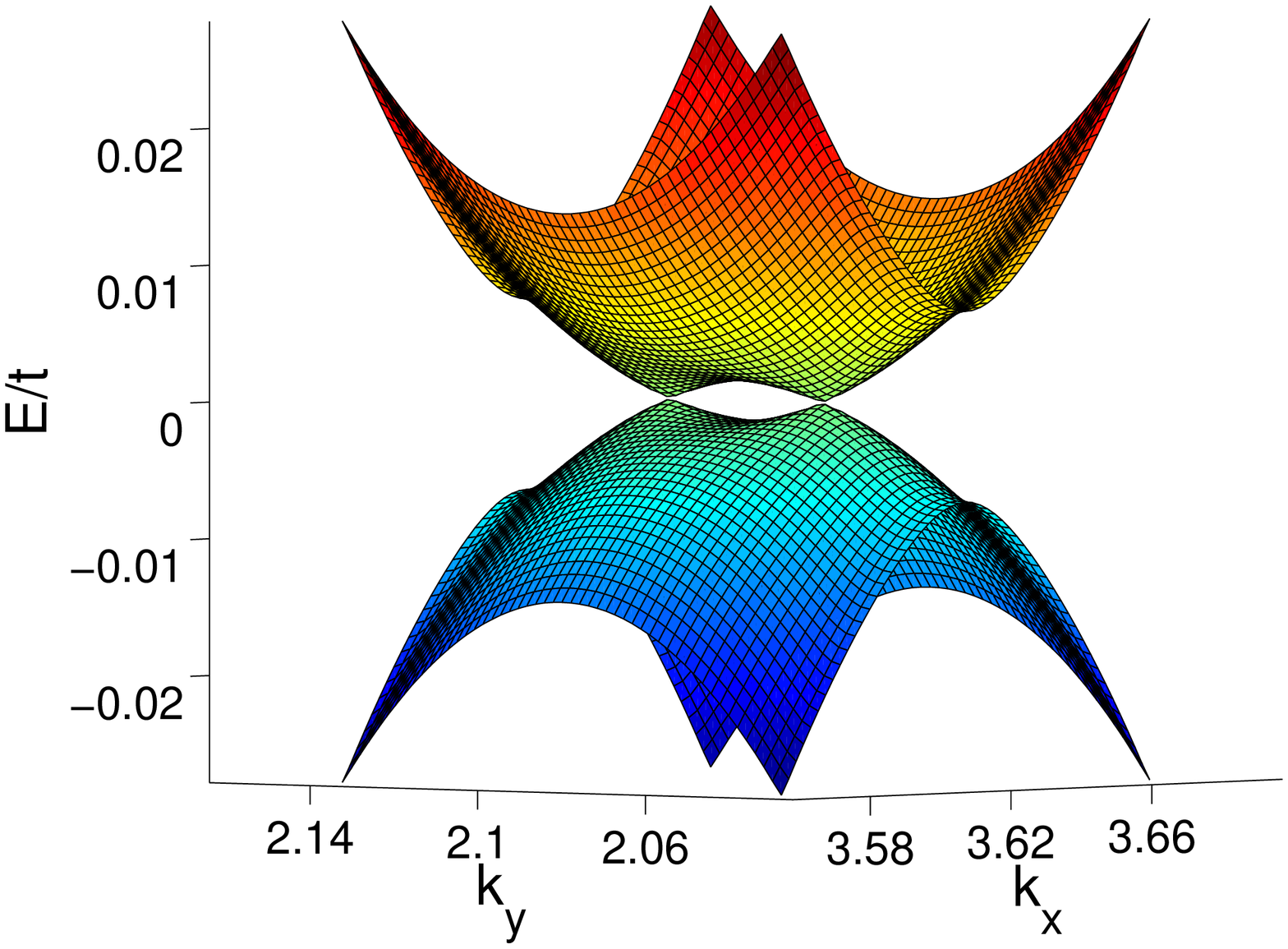} 
	\includegraphics[width=.22 
	\textwidth]{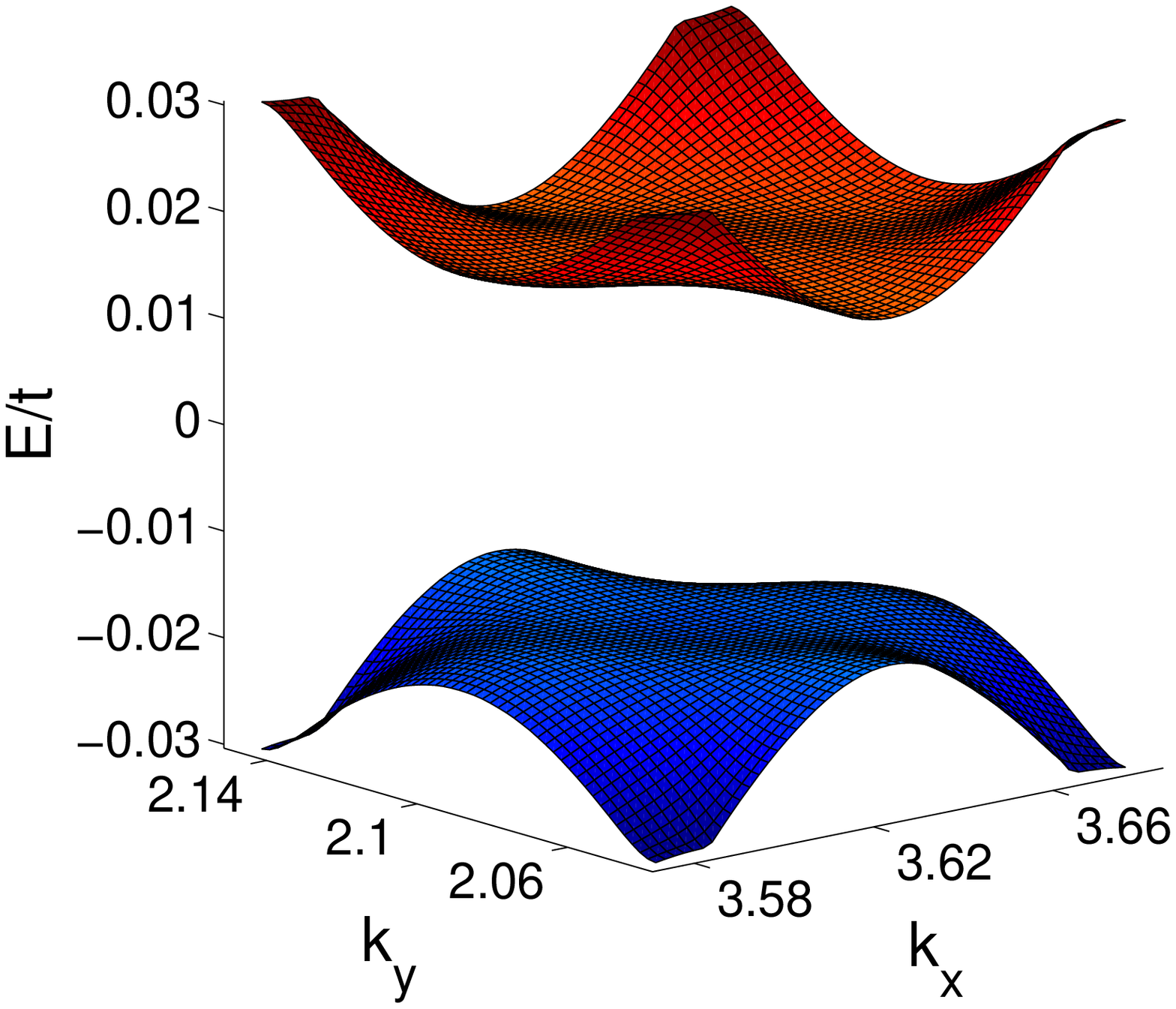} \caption{Energy spectrum for a 1D superlattice with step-like chemical potential modulation of amplitude $U$. We set $\lambda=60d$, with [left panel] $U=0.01t$ showing two Dirac nodes split along $\hat{y}$ near $\bK$, and with [right panel] $U=0.04t$ showing a full gap.} \label{fig1} 
\end{figure}

{\it 1D chemical potential superlattice. ---} Imposing a periodic potential $V_{1}(x,y)=V_2(x,y)=U(x,y)$ corresponds to a chemical potential modulation. Numerically solving for the band structure of a periodic 1D modulation using the above effective Hamiltonian, we find a pair of zero energy Dirac points in the MBZ in the vicinity of each valley. This is shown in Fig.\ref{fig1} for a periodic step-like potential with (i) $U(x,y)=U$ for $0 \leq y < \lambda/2$ and (ii) $U(x,y)=-U$ for $\lambda/2 \leq y < \lambda$. With increasing $U$, these Dirac points move away from each other along $\hat{y}$. Beyond a critical modulation amplitude a full gap opens up.

The existence of two Dirac cones at each valley is deeply rooted in the chiral nature of the low energy BLG quasiparticles, which causes the matrix elements of Eqn.~\ref{sl} to depend on the scattering angle $\theta$. For states with momenta parallel to the modulation direction, $\theta=0$ or $\pi$, the off-diagonal matrix elements vanish; the electron and hole states then decouple, but electron-electron and hole-hole mixing is allowed. However, in an extended zone scheme, all such electron (hole) states within the first MBZ only mix with electron (hole) states of higher (lower) energy, and so the energy of these states will be globally shifted down (up). This results in two level crossings along the modulation direction, which are protected by the chirality of the low energy BLG quasiparticles. If this electron-hole decoupling was true for all momenta, we would see the two parabolic bands crossing on a full circle in the MBZ, but going to momenta $(\delta p_x, p_y)$ leads to electron-hole mixing that is linear in $\delta p_x$; this results in an avoided level crossing and the robust emergence of two Dirac cones in the MBZ.

The location and velocity anisotropy of Dirac cones, as well as the critical modulation amplitude to gap them out, can be predicted using perturbation theory in $U(\bG)$. The second order energy correction of states with ${\bf p}=(0,p_y)$ is $ \Delta E^{(2)}({\bf p})=\sum_{n\neq 0}{|U(n{\bf Q})|^2}/\left[ {\varepsilon_{e,h}({\bf p})-\varepsilon_{e,h}({\bf p}+n{\bf Q})}\right]. $ Since ${\varepsilon_{e}({\bf p})<\varepsilon_{e}({\bf p}+n{\bf Q})}$ while ${\varepsilon_{h}({\bf p})>\varepsilon_{h}({\bf p}+n{\bf Q})}$ in the MBZ, this correction is always negative (positive) for electron (hole) states, as expected.

Thus, the two bands will intersect and cross linearly at momenta $(0,\pm p^*_y)$, where $ {p^{*2}_y}/{2m^*}=2m^* \sum_{n \neq 0} |U(n{\bf Q})|^2 /\left[{n^2Q^2+2p^*_y nQ}\right]. $ For weak modulations, $p^*_y/Q \! \ll\! 1$, and keeping only $n\!=\!\pm 1$, we estimate $p^*_y \! \approx \! \sqrt{2}m^* |U(\bQ)| \lambda/\pi$. For a step profile, $|U(\bQ)| \!=\! 2 U/\pi$, and $|n| \! > \! 1$ contributions are small. 

For small $\delta p_x$ away from the level crossing point, we can estimate the electron-hole mixing term using perturbation theory \cite{Suppl}, and we find that the resulting eigenstates have energies $\epsilon_\bp= \pm (16 m^* |U(\bQ)|^2 / |\bQ|^2) \delta p_x/p^*_y$. The crossing points at $(0,\pm p^*_y)$ are thus really massless Dirac points in the full MBZ. We find velocities $v_y = p^*_y/m^* \approx \sqrt{2} \lambda |U(\bQ)|/\pi$, and $v_x = 2 v_y$ for the anisotropic linear dispersion.

Once these Dirac nodes reach the MBZ boundary, Bragg scattering between them opens up a full gap. The critical potential strength, $|U_c(\bQ)|$ for this is roughly estimated by setting $p_y^*=Q/2$, which yields $|U_c(\bQ)| \approx \pi^2/(\sqrt{2}m^*\lambda^2)$. For a step profile, with $\lambda=60d$, we find $U_c \approx 0.03 t$ which is close to the numerical result $0.02t$. 

\begin{figure}
	[t] \centering 
	\includegraphics[width=.23 
	\textwidth]{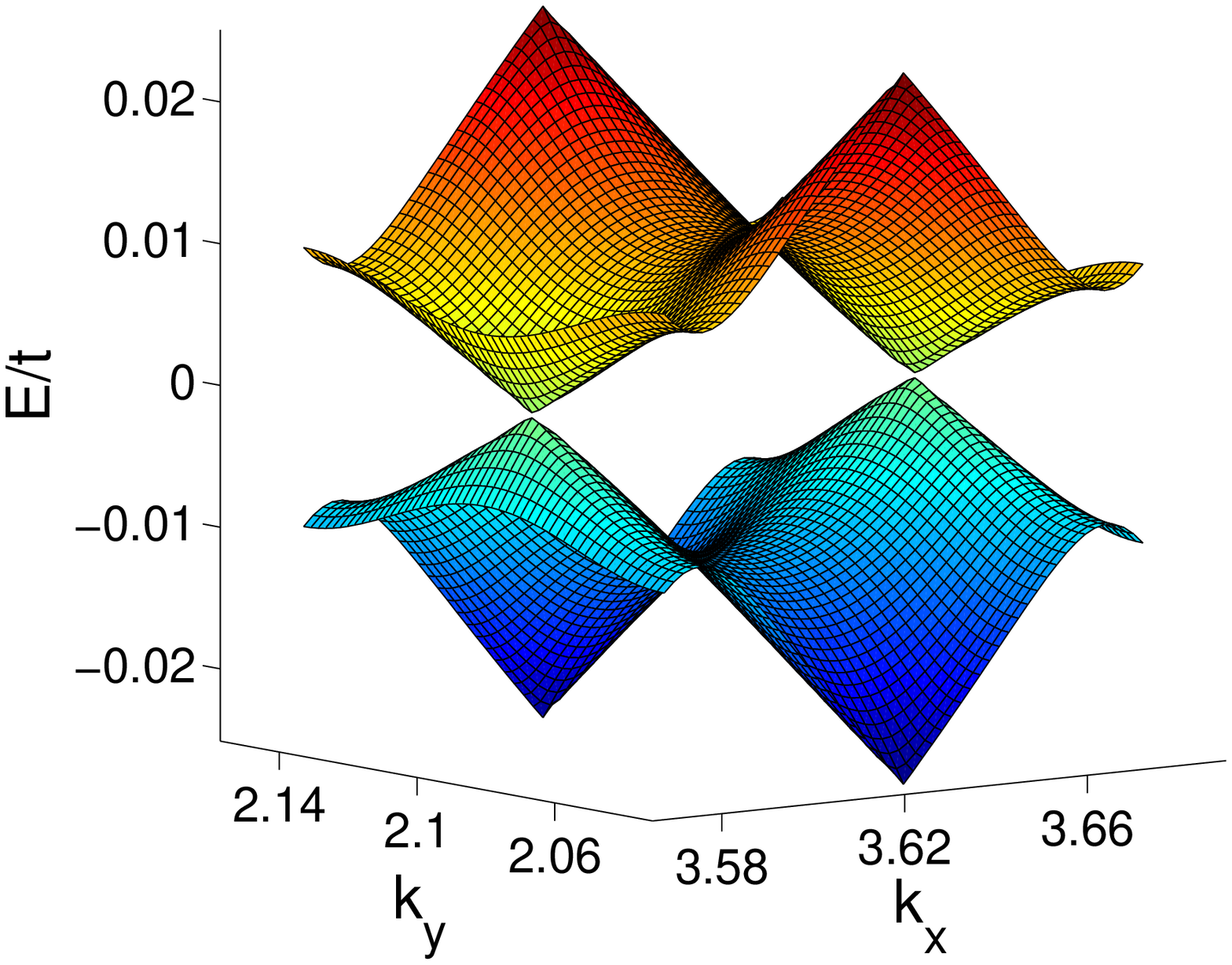} 
	\includegraphics[width=.23 
	\textwidth]{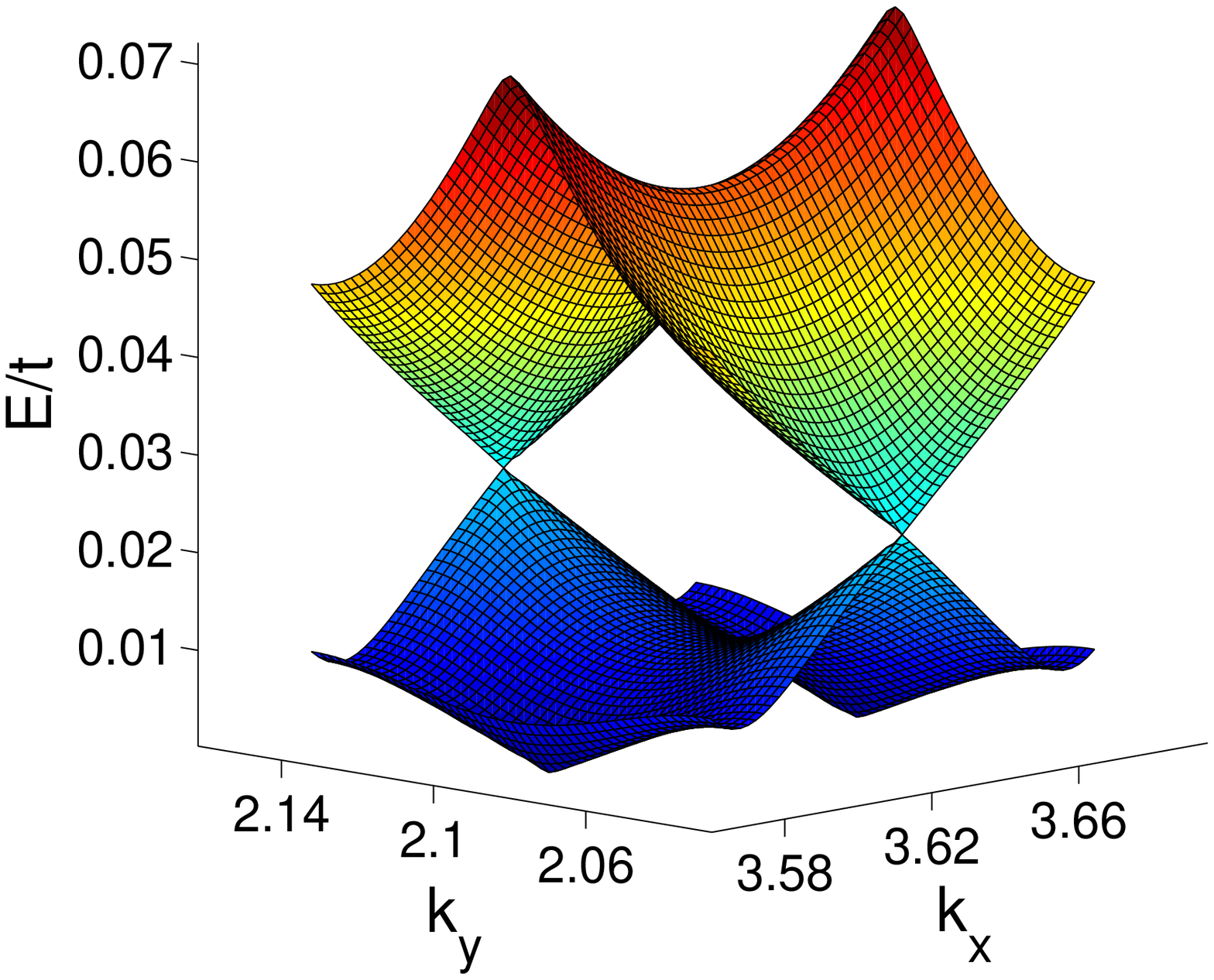} \caption{Energy spectrum for a 1D {\it symmetric} (see text) electric field superlattice with $\lambda=60d$ and $U=0.03 t$, showing a pair of zero energy massless Dirac fermions at $(\pm p^*_x,0)$ [left panel] and a nonzero energy Dirac point at $(0,\pm \pi/\lambda)$ [right panel].} \label{fig3} 
\end{figure}

{\it 1D electric field superlattice.---} An electric field SL corresponds to $V_1(x,y)=-V_2(x,y)=U(x,y)$. Solving for the resulting band structure, we find that it depends sensitively on the modulation type. To illustrate this, we consider a periodic potential, with $U(y)=2 U (1-w/\lambda)$ for $0 \leq y < w$, and $U(y)=-2 U w/\lambda$ for $w \leq y < \lambda$. We have set the average potential on each layer to be zero. If $w=\lambda/2$, the resulting {\it symmetric} SL is found to support a pair of anisotropically dispersing massless Dirac fermions at zero energy at $(\pm p^*_x,0)$, as seen in Fig.~\ref{fig3} (left panel). In addition, as shown in Fig.~\ref{fig3} (right panel), it supports a Dirac point at nonzero positive (as well as negative) energies at $(0,\pi/\lambda)$ (or equivalently $(0,-\pi/\lambda)$). However, an {\it asymmetric} SL, with $w \neq \lambda/2$, leads to a gap for all these Dirac fermions. More generally, we find that if the SL potential commutes with a generalized parity operator, ${\cal P}$, which corresponds to $y \to -y$ followed by exchanging the two layers of BLG, then these gapless Dirac points survive. Breaking ${\cal P}$ leads to gaps.

\begin{figure}
	[b] 
	\includegraphics[height=2.8cm]{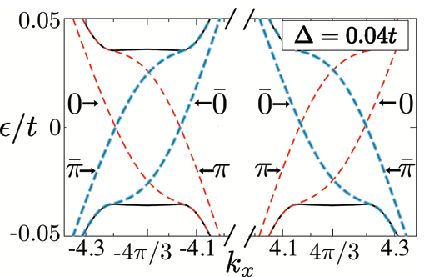} \quad \quad 
	\includegraphics[height=2.8cm]{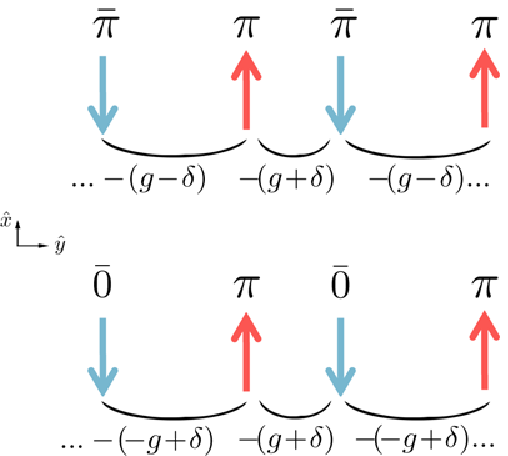} \caption{(color online) Left: Spectrum of isolated kink (thin, red) and antikink (thick, blue). Higher (lower) energy modes are labelled $\pi$ ($0$) at a kink and as $\bar{\pi}$ ($\bar{0}$) at an antikink. Right: Schematic of hopping between the $\pi-\bar{\pi}$ and $\bar{0}-\pi$ states.} \label{wiredispersion} 
\end{figure}

A simple route to understanding these results that leads to other interesting predictions is to view the SL as a periodic array of `kinks' and `antikinks' where a kink (antikink) corresponds to where the electric field flips from pointing up (down) to pointing down (up). A single such kink/antikink in the bias is well understood \cite{Martin,Killi,Xavier,LiJian}. In the absence of interactions a kink (antikink) supports a pair of right-moving (left-moving) `topological' edge states near the ${\bf K}$ point for each spin. By time-reversal, these right and left movers get interchanged at the $-{\bf K}$ point. These modes are depicted in Fig.~\ref{wiredispersion}. (Although these modes were suggested to be topologically protected, they are not truly stable against disorder; nevertheless disorder induced backscattering is weak \cite{LiJian}.) At a kink, we denote the higher (lower) energy edge state as $\pi$ ($0$), while we denote these states as $\bar{\pi}$ ($\bar{0}$) at an antikink. Hence, there are four points at each valley where kink and antikink modes cross: two of these occur at zero energy ($\pi$-$\bar{0}$ and $\bar{\pi}$-$0$ crossings), and two of them occur at nonzero energy ($\pi$-$\bar{\pi}$ and $0$-$\bar{0}$ crossings). We will show below that these crossing points evolve into massless Dirac fermion modes in the MBZ of the SL. In order to see this, we construct a tight-binding model of such coupled `topological' edge states. 

We observe that the Hamiltonian with the single kink (or antikink) potential is invariant under ${\cal P}$, since $\mathcal{P^{\dag}}H(y)\mathcal{P}\!=\!\sigma_x H(-y) \sigma_x \!=\! H(y)$. The $0/\bar{0}$ states are {\it even} under ${\cal P}$, while the $\pi/\bar{\pi}$ states are {\it odd} under ${\cal P}$ \cite{Martin}. Let us then construct a reduced Hamiltonian which describes the hybridization between neighboring edge modes.

We begin with neighboring $\pi$-$\bar{0}$ modes at zero energy and at a momentum $p^*_x$ (away from ${\bf K}$). The hopping between neighboring `wires' along $\hat{y}$ is then between states which have opposite velocities (since it is between a kink and an antikink edge state) and it is between a p-wave like state (${\cal P}$-odd) and an s-wave like state (${\cal P}$-even). Using the index $n$ to label the wires, the interchain hopping parameter will then alternate as $(-1)^n g$ for equally spaced wires and as $g+\delta, -g+\delta$ (with $\delta < g$) if pairs of wires are closer to each other \cite{Suppl}. Linearizing the dispersion at the crossing point, and letting $v_0$ denote the velocity of the linearized modes, 
\bea H(\px)=&v_0&\sum_n \left((-1)^n (p_x-p^*_x) c^{\dg}_{\px n}c_{\px n}\right) \notag \\
&-&\sum_n (g (-1)^n + \delta) \left(c^{\dg}_{\px n}c_{\px n+1} + h.c.\right) \eea 
where $p^*_x$ is the location of the $\pi-{\bar 0}$ crossing point in the single kink or antikink problem, and $c_{\px n}$ annihilates an electron on wire $n$ with momentum $\px$. Let $\xi(p_x)\equiv v_0 (p - p^*_x)$. Fourier transforming, we find $H(\px)=\sum'_\py \Psi^\dg(\py) {\bf \sigma} \cdot {\bf h}(\px) \Psi(\py)$, where ${\bf h}(\px)=\left(\xi(p_x) , -2 g \sin(\py), -2 \delta \cos(\py)\right)$, with $\Psi(\py) =(c_\py \, c_{\py+\pi})^T$, and $\sum'_\py$ runs over the MBZ. The dispersion is thus $E=\pm \sqrt{\xi^2(p_x) + 4 \delta^2 \cos^2(\py) + 4 g^2 \sin^2(\py)}$. Consequently, when $w=\lambda/2$, and the Hamiltonian commutes with ${\cal P}$, we have $\delta=0$ and a Dirac cone is generated at $(p^*_x, 0)$, consistent with numerical results. When $w \neq \lambda/2$, the Hamiltonian breaks ${\cal P}$ --- we then have $\delta \neq 0$, which leads to a gap $4\delta$. Similar arguments hold for the other zero energy band crossing points. The velocity of the Dirac fermions is highly anisotropic and depends on $g$ --- this can be controlled by tuning the SL period and amplitude.

The above analysis can also be repeated for the nonzero energy ($0$-$\bar{0}$ and $\pi$-$\bar{\pi}$) crossings \cite{Suppl}; in the {\it symmetric} case, $w=\lambda/2$, we find Dirac cones at $(0,\pm \pi/\lambda)$ on the MBZ. Once again, a modulation with $w\neq\lambda/2$ results in a finite $\delta$ and opening of band gap.

Interestingly, just as in polyacetylene, a domain wall between a gapped region with $w >\lambda/2$ and a gapped region with $w <\lambda/2$ leads to new subgap soliton modes. Since each kink/antikink is itself like a domain wall, these should be viewed as solitons in a soliton lattice!

{\it 2D superlattices.---} We have also considered 2D chessboard like SLs with fourfold rotation symmetry. For both types of 2D SLs, chemical potential or electric field, the quadratic band touching point remains intact when the SL potential is `symmetric', $ V_{1,2}(x+\lambda/2,y)=V_{1,2}(x,y+\lambda/2)=-V_{1,2}(x,y). $ This is consistent with the fact that no Dirac points can be generated in a way that conserves both topological charge and $C_4$ (or $C_6$) symmetry \cite{Sun}. For asymmetric SLs, higher order corrections lead to modifications to the energy spectrum at the {\bf K}-point \cite{Suppl}. For chemical potential SL, the charge neutrality point (CNP) shifts slightly in energy, due to higher order effects which reflect particle-hole symmetry breaking. For electric field SLs, breaking generalized parity opens a small gap at the ${\bf K}$-point \cite{Suppl}. 

{\it Experimental implications.---} Our work demonstrates that SL modulations in BLG can generate new Dirac fermion modes. Such modes are perturbatively stable to interaction effects, and could be experimentally explored by suitable choice of substrates. Disorder will also lead to such bias and chemical potential modulations, albeit in random fashion. One source of such fluctuations is the presence of charged impurities, embedded in the underlying substrate (SiO$_2$) or, in the case of suspended BLG, in the residue of the etching/washing process. Such impurities are expected to locally shift the CNP, and to suppress or enhance the bandgap depending on the relative sign of the bias and the impurity electric field \cite{Deshpande}. If the impurity lies close to the surface it can locally reverse the parity of the interlayer bias leading to `topological' subgap modes. Another source of SL fluctuations is rippling \cite{Ishigami, Bao}, which would modulate the electric field perpendicular to the bilayer at the ripple wavelength. 
\begin{figure}
	[t] \centering 
	\includegraphics[height=3.1cm]{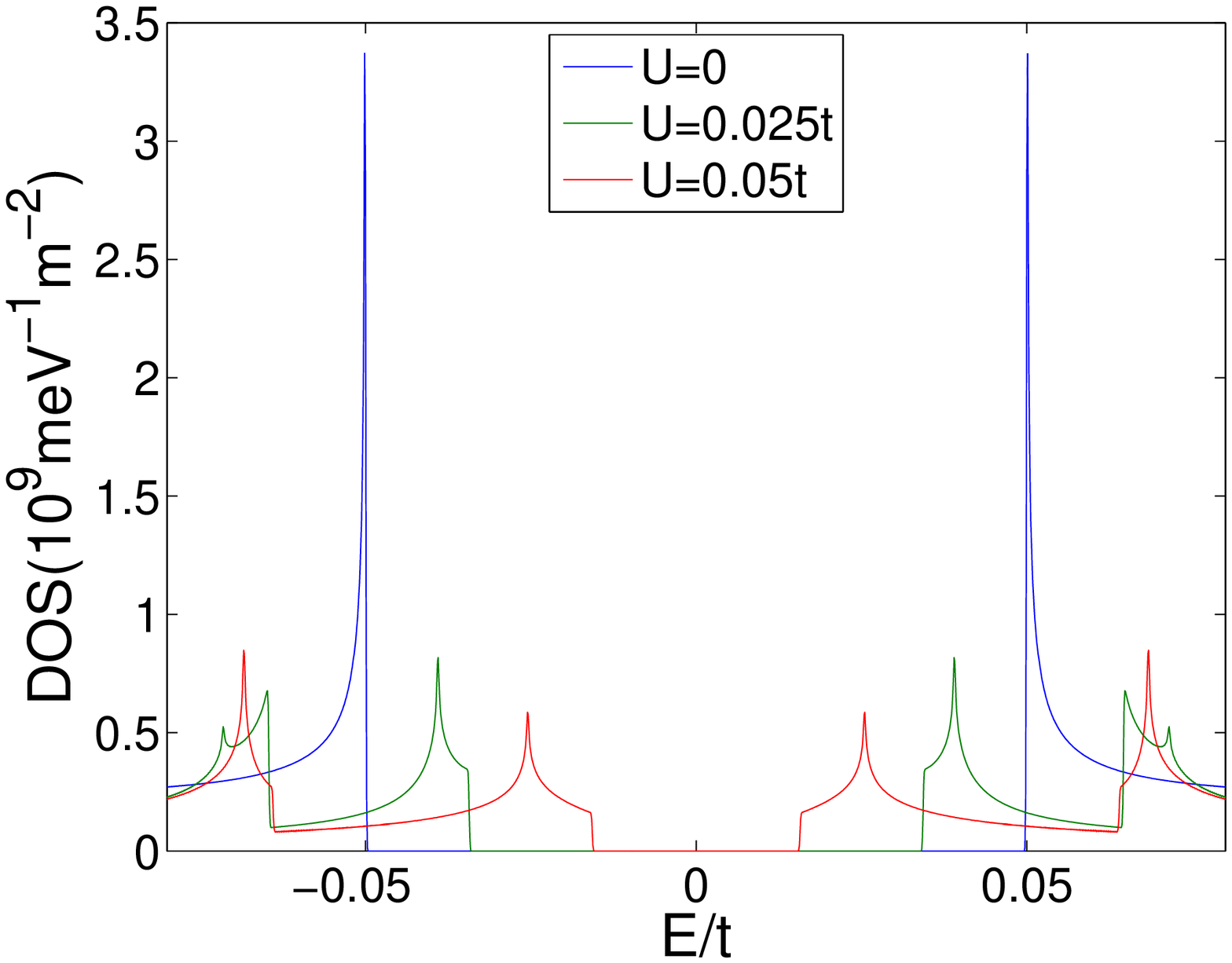} \quad 
	\includegraphics[height=3.1cm]{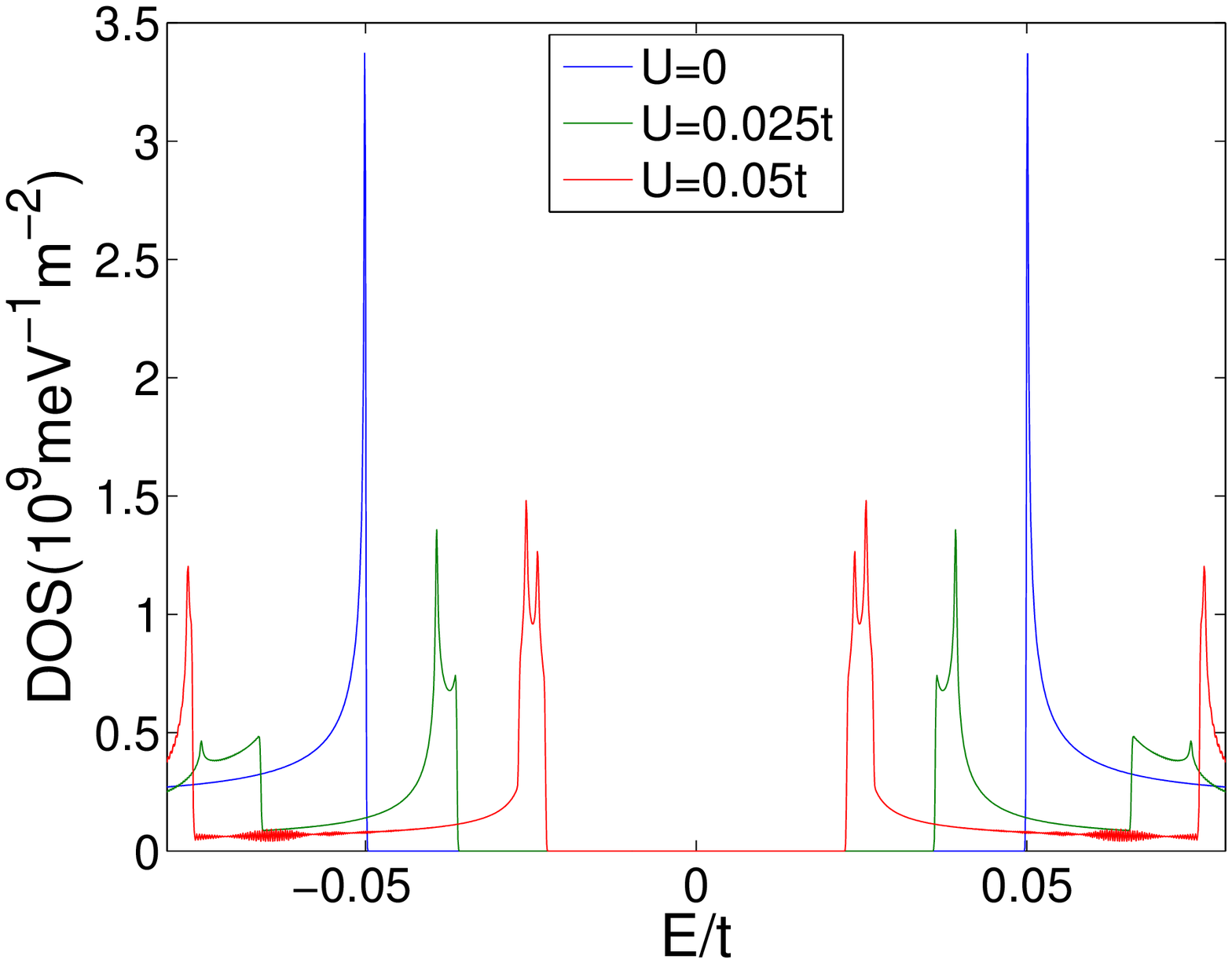} \caption{(color online) Density of states for BLG subject to a uniform bias of $\Delta=0.1t$ and various chemical potential (left) and electric field (right) superlattices with period $\lambda=60d$.} \label{fig4} 
\end{figure}

As a starting point to understanding the expected role of chemical potential and electric field fluctuations, Fig.\ \ref{fig4} shows density of states (DOS) plots of a biased SL with periodic 1D modulations. In the absence of a SL, the DOS diverges as $1/\sqrt{E}$ at the gap edge arising from the $\sim p^4$ dispersion of modes near the gap edge. We find that both chemical potential or bias modulations, cause low energy subgap modes states in this system that will renormalize the average band gap, consistent with experiment. For chemical potential modulations, the subgap states are due to the local shift in the CNP. At finite temperature, regions with a slightly shifted CNP will have thermally activated `electron-hole' puddles that contribute to transport. For bias modulations, weak modulations locally enhance or suppress the bandgap, while strong modulations form `topological' states in the bulk along interfaces where the field reverses sign \cite{Martin,Killi, Xavier,LiJian}. The energy of these `topological' midgap states $decreases$ for large and dilute fluctuations, as the overlap between edge mode wavefunctions is reduced.

{\it Random} potential fluctuations will have two important effects not captured in our study of periodic modulations. First, it will cause the low energy density of states to broaden, causing further suppression of the bandgap predicted by the periodic modulation. Second, dilute localized `topological' states induced in the bulk by strong random electric field modulations due to charged impurities will contribute to transport through variable range hopping --- this is broadly consistent with the temperature dependence of the resistance in biased BLG \cite{Oostinga,Tay,Miyazaki,Yan}.

This work was supported by NSERC, an Ontario ERA, and the Indian DST. MK and AP acknowledge the hospitality of ICTS-TIFR (Bangalore).

{\it Note Added:} After submission of this Letter, we received a preprint of Ref.~\cite{Tan}, which studies Dirac fermions in 1D chemical potential superlattices in BLG and contains results consistent with ours.

\end{document}


\title{Supplemental Material for ``Band structures of bilayer graphene superlattices"}
\author{
Matthew Killi}
\affiliation{Department of Physics, University of Toronto, Toronto, Ontario, 
Canada M5S 1A7}
\author{Si Wu}
\affiliation{Department of Physics, University of Toronto, Toronto, Ontario, 
Canada M5S 1A7}  
\author{Arun Paramekanti}
\affiliation{Department of Physics, University of Toronto, Toronto, Ontario, 
Canada M5S 1A7} 
\affiliation{Canadian Institute for Advanced Research, Toronto, Ontario, M5G 1Z8, Canada}
\affiliation{Department of Physics, Indian Institute of Science, Bangalore, India 560 012}

\maketitle

\section{1D  Chemical Potential Superlattice: Perturbation Theory}
\textit{Location of the Dirac Point} ---
For small bias strengths, the location of Dirac point, $(0,p^*_y)$, can be estimated by setting the absolute value of the second order energy correction equal to the free electron energy,
\bea	\label{E}
	\frac{p^{*2}_y}{2m^*}=2m^* \sum_n|U(n{\bf Q})|^2 \! &\!&\! \! \! \! \left(\frac{1}{n^2Q^2+2p^*_y nQ}\right. \notag\\
		&&+\left. \frac{1}{n^2Q^2-2p^{*}_y nQ}\right).
\eea
Equation \ref{E} can then be expanded in the $p_y/Q \ll 1$ limit to give

\bea	\label{E2}
	p^{*2}_y \approx \frac{2m^{*2}\lambda^{2}}{\pi^{2}} \sum_n \! &\!&\! \! \! \!\frac{|U(n{\bf Q})|^2}{n^2},
\eea
or retaining only the fundamental harmonic

\bea \label{E3}
	p^*_y \approx \frac{\sqrt{2}m^{*} |U({\bf Q})| \lambda}{\pi}.
\eea

For the case of a square potential where $|U(n \bQ)|=2U/n \pi$ for odd $n$ and keeping only the fundamental harmonic,
 \bea \label{p}
 p^*_y \approx 2\sqrt{2}m^{*} U \lambda / \pi^{2}.
 \eea
 Alternatively, Eq.\ \ref{E2} can be computed explicitly through the 
relation $\sum_{n_{odd}}1/n^4=\pi^4/96$ to give $p^*_y \approx \pm \frac{\sqrt{3}m^*U \lambda}{6}$.  A comparison of the two expressions confirms that higher order harmonic corrections are indeed small $(\sim 1\%)$ and can be ignored.

\textit{Critical Superlattice Strength} ---
Using the approximation provided for $p^{*}_{y}$ in Eq.\ \ref{E3}, it is also possible to make a crude estimate of the critical potential, $U_c$, before the onset of a bandgap.  From numerical results consistent with the perturbative results above, the Dirac points are seen to move out towards the mini-zone boundary with increasing $U$.  Upon reaching the mini-zone boundary, any further increase in $U$ results in an opening of a bandgap at the mini-zone boundary.  Hence, the critical potential strength, $U_c$, can be determined from the condition that the Dirac point $p^*_y=p_c=\pm Q/2$.  Although the above perturbative result is not strictly valid in this regime, it provides a crude estimate of
\bea
	|U_c (\bQ)| \approx \frac{\pi^2}{\sqrt{2}m^*\lambda^2}.
\eea

In the case of a step potential with $\lambda=60d$, $U_{c}\approx \frac{\pi^3}{2\sqrt{2}m^*\lambda^2} \approx 0.03t $, which is reasonably close to the observed numerical value of $0.02t$.

\textit{Velocity Anisotropy} --- 
Also of interest is the degree of anisotropy of the group velocity about the Dirac cone.  Along the $\bp=(0,p_y)$ direction, the above perturbation theory indicates that $v_y=p^*_y/m^* \approx\frac{\sqrt{2}\lambda |U({\bf Q})|}{\pi}$.  To calculate the velocity along the $p_x$ direction, we perform a degenerate perturbation theory for an electron and hole states with momentum $\bp=( p^*_y \, \theta, p^*_y)$ for a small angle $\theta$ while retaining only the leading order harmonic.  For such states, the off-diagonal terms in the Hamiltonian reduce to,
\bea
	W_{\bp, \bp-\bQ}=
	\left(
	\begin{array}{cc}
	U(\bQ) & \frac{iU(\bQ)}{2}\sin (2 \theta_{\bp, \bp-\bQ}) \\
	\frac{iU(\bQ)}{2}\sin(2\theta_{\bp, \bp-\bQ}) & U(\bQ)
	\end{array}
	\right).
\eea

Given the small finite momentum along the $p_x$ direction, the originally pure electron states get mixed with hole states and vice versa.  To leading order in $\theta$,
\bea
	|\Psi^e_\bp \rangle =|\alpha_\bp \rangle 
	&+&\frac{U(\bQ)|\alpha_{\bp+\bQ} \rangle}{\epsilon_e(\bp)-\epsilon_e(\bp+\bQ)} 
	+\frac{U^*(\bQ) |\alpha_{\bp-\bQ} \rangle}{\epsilon_e(\bp)-\epsilon_e(\bp-\bQ)} \nonumber \\ 
	&-&\frac{i\theta U(\bQ)  |\beta_{\bp+\bQ}\rangle}{\epsilon_e(\bp)-\epsilon_h(\bp+\bQ)} 
	- \frac{i\theta U^*(\bQ) |\beta_{\bp-\bQ} \rangle}{\epsilon_e(\bp)-\epsilon_h(\bp-\bQ)} \notag
\eea
and,
\bea
	|\Psi^h_\bp \rangle =|\beta_\bp \rangle 
	&+&\frac{U(\bQ)|\beta_{\bp+\bQ} \rangle}{\epsilon_h(\bp)-\epsilon_h(\bp+\bQ)} 
	+\frac{U^{*}(\bQ) |\beta_{\bp-\bQ} \rangle}{\epsilon_h(\bp)-\epsilon_h(\bp-\bQ)} \nonumber \\ 
	&-&\frac{i\theta U(\bQ)  |\alpha_{\bp+\bQ}\rangle}{\epsilon_h(\bp)-\epsilon_e(\bp+\bQ)} 
	- \frac{i\theta U^*(\bQ) |\alpha_{\bp-\bQ} \rangle}{\epsilon_h(\bp)-\epsilon_e(\bp-\bQ)}. \notag
\eea
Since, $\langle\Psi^e_\bp | H |\Psi^e_\bp \rangle = \langle\Psi^h_\bp | H |\Psi^h_\bp \rangle=0$ to first order in $\theta$ and are degenerate, we must compute the off-diagonal matrix elements $\langle\Psi^e_\bp | H |\Psi^h_\bp \rangle$ of the Hamiltonian.  To leading order in $\theta$,
\bea
	\langle\Psi^e_\bp | H |\Psi^h_\bp \rangle \!&=\!&-2i\theta |U(\bQ)|^2 
	\left( \frac{1}{\epsilon(\bp)-\epsilon(\bp+\bQ)}\right. \notag \\
	&+&\frac{1}{\epsilon(\bp)-\epsilon(\bp-\bQ)}
	  +\frac{1}{\epsilon(\bp)+\epsilon(\bp+\bQ)} \notag \\
	&+&\!\left. \frac{1}{\epsilon(\bp)+\epsilon(\bp-\bQ)}\right),
\eea
where $\epsilon(\bp)\equiv\epsilon_e(\bp)=-\epsilon_h(\bp)$.  In the small $\bp$ limit, the matrix element reduces to
\bea
	\langle\Psi^e_\bp | H |\Psi^h_\bp \rangle=-\frac{16i m^* \theta |U(\bQ)|^2 }{|\bQ|^2}.
\eea
This give the solution $\epsilon_\bp= \pm 16 m^* \theta |U(\bQ)|^2 / |\bQ|^2$ to the perturbed Hamiltonian is then, from which the velocity can be calculated.  Using the relation $p^*_y \, \theta=p_x$ for small the theta,
\bea
	\epsilon(p_x)&=&\frac{16m^*|U(\bQ)|^2 p_x}{p^*_y \, |\bQ|^2} \nonumber \\
	\rightarrow &v_x&=2\sqrt{2}\lambda |U({\bf Q})|/\pi
\eea
Hence, the anisotropy of the velocity in the Dirac cone is predicted to be $v_x/v_y = 2$ for small U, which is again remarkably consistent with the numerical results.

The above results are based on the two-band reduced Hamiltonian for BLG. We have also carried out a similar 
calculation for the four-band Hamiltonian. We found the same behavior when superlattice potential is not 
very strong. Surprisingly, when the superlattice potential is comparable to $t_{\perp}$, more band touching 
points emerge in the mini Brillouin zone.

\section{1D Electric Field Superlattice: Effective Model}
In this section, we demonstrate in detail how the Dirac cones generated by a symmetric bandgap modulation ($w=\lambda/2$) can be understood in terms of a tight-binding theory that describes a chain of coupled 1D wires.

\textit{Transfer Integrals} ---
To derive the relation between the transfer integrals connecting wire $n$ and $n+1$, and $n-1$ and $n$, careful consideration of the symmetry of the soliton wavefunctions must be made.   As stated in the main text, the principle symmetry of the wavefunctions follow from the invariance of the Hamiltonian under the combined operation of layer inversion and reflection about a kink (or antikink), i.e.\ $\mathcal{P^{\dag}}H(y)\mathcal{P}=\sigma_x H(-y) \sigma_x=H(y)$.  Solutions are then of the form
\bea
	\left( 
	\begin{array}{cc}
	f(y) \\ g(y)
	\end{array}
	\right)=\left(
	\begin{array}{cc}
	f(y) \\ f(-y)
	\end{array}
	\right),\left(
	\begin{array}{cc}
	f(y) \\ -f(-y)
	\end{array}
	\right),
\eea
with corresponding eigenvalues of $+1$ and $-1$ of the operator $\mathcal{P}$, respectively.  For the case of the zero energy band crossing points (between the $\pi$- and $\bar{0}$-modes or the $\bar{\pi}$- and $0$-modes at either K-point), the soliton wavefunctions of the kink wire have $opposite$ $\mathcal{P}$ symmetry to the soliton wavefunctions of the two neighbouring anti-kink wires.  In contrast, for the case of the finite band crossing points (between the $\pi$- and $\bar{\pi}$-modes or the $\bar{0}$- and $0$-modes at either K-point), the soliton wavefunctions of the kink wire have the $same$ $\mathcal{P}$ symmetry as the soliton wavefunctions localized to its two neighbouring anti-kink wires.  As we will now show, the transfer integrals describing the hopping between neighbouring wires along the array is dependent on whether the parity of the coupled modes is the same or opposite.

For concreteness, let us consider the region where the $\pi$- and $\bar{0}$-bands cross at the K point for a symmetric modulation with $w=\lambda/2$.  Let us set the $y=0$ point to be an anti-kink wire.  The anti-kink $\bar{0}$-modes have wavefunctions of the form $\Psi^{\bar{0}}(y)=(w(y),w(-y))^T$ while its two neighbours' $\pi$-modes have wavefunctions of the form $\Psi^\pi(y+\lambda/2)=\left(v(y+\lambda/2),-v(-y-\lambda/2)\right)^T$ and $\Psi^\pi(y-\lambda/2)=\left(v(y-\lambda/2),-v(-y+\lambda/2)\right)^T$, respectively.  For simplicity, we will assume that the wavefunction overlap is finite only in the region between the wires, as this assumption does not effect our main result.  The transfer matrix that determines the hopping parameter between the central wire and its left neighbour is then
\bea
	g_1=\int^0_{-\lambda/2}  \Psi^{\bar{0} \dag}(y) \, H(y) \, \Psi^\pi(y+\lambda/2) \,  \mathrm{d}y,
 \eea
 and between its right neighbour
 \bea
	g_2=\int^{\lambda/2}_{0}  \Psi^{\pi \dag}(y-\lambda/2) \, H(y) \, \Psi^{\bar{0}}(y) \,  \mathrm{d}y.
 \eea

Inserting $H(y)= \sigma_x H(-y) \sigma_x$ and changing $y \rightarrow -y$ in the expression for $g_2$, gives us the relation
\bea
	g_1=-g_2^*\equiv |g|e^{i \theta}
\eea
between alternate bonds.  The same relationship holds for the other zero energy band crossing point.  However, repeating this calculation for the finite energy band crossing points (where the wavefunctions have the same parity) yields the corresponding relation
\bea
	g_1'=g_2^{*}{'}\equiv t=|g'|e^{i \theta} .
\eea

Without loss of generality, in both cases the hopping parameter be assumed to be real, as it is always possible to remove the phase factor by a simple gauge transformation.  Hence, we have deduced from very general symmetry arguments that when $w=\lambda/2$ the hopping parameter between the $0$- ($\bar{0}$-) and $\pi$- ($\bar{\pi}$-) modes of neighbouring wires alternates sign along the chain, and is uniform between $0$- ($\pi$-) and $\bar{0}$- ($\bar{\pi}$-) modes of neighbouring wires.

If we generalize this case where $w\neq\lambda/2$, the separation between neighbouring wires is unequal and the magnitude of the hopping parameter will begin to alternate along the bonds.  Again considering the $\pi$- $\bar{0}$-band crossing at the K point and taking the the wire at larger $y$ to be further than to the anti-kink wire than the other neighbour, $g_{1}=-\left(-g+\delta\right)$ and $g_{2}=-\left(g+\delta\right)$, where $g>0$ is the average magnitude of the hopping between neighbouring wires and $\delta>0$ is the deviation.  Alternatively, for the finite energy band crossing points, the hopping parameter can be shown to be $g_{1}=-\left(g-\delta\right)$ and $g_{2}=-\left(g+\delta\right)$.

\begin{figure}[t]
	\centering
	\includegraphics[width=0.44\textwidth]{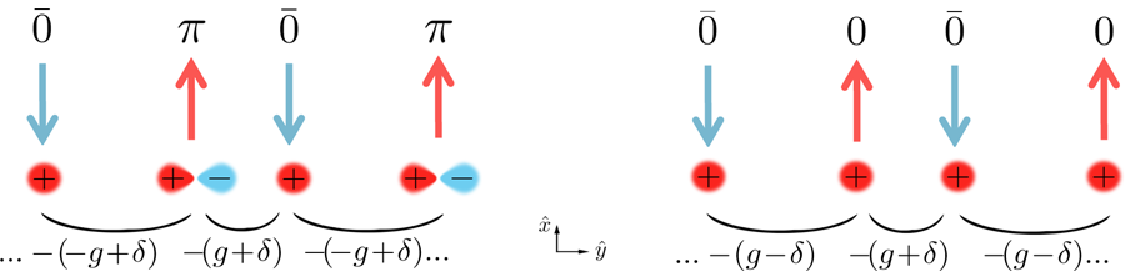}
	\caption{(Color online) Variation of the hopping parameter along the wire array.  Left:  Hopping between zero energy modes of opposite parity.  Right:  Hopping between finite energy modes with the same parity.   Shape, orientation and sign of the wavefunctions are completely schematic and serve only to be illustrative of effect of parity on the hopping integrals.}
\end{figure}
\textit{Velocity Anisotropy} ---
Along the wire direction, $\bp=(\px,0)$, the group velocity is the same as that of a single kink anti-kink pair at the band crossing point.  This velocity is only sensitive the details of the bias profile within the unit cell, and independent of the modulation period.  In contrast, along the modulation direction, $\bp=(0,\py)$, the velocity is given by the effective interwire hopping strength, $g$, and is thus dependent on the period of the modulation, as $g$ is given by the overlap between the wires.  As a consequence, it is interesting that the velocity along the each direction can be tuned independently.  After tuning the velocity along the $\bp=(\px,0)$ direction, the velocity along the $\bp=(0,\py)$ direction can be tuned continuously by adjusting the period length while keeping the velocity along the $\bp=(\px,0)$ direction fixed.

\textit{Limits of Applicability} ---
After having analyzed the low-energy model, we now comment on the limits in which this model can be applied.  As with other tight-binding models, its validity is dependent on the extent of overlap between adjacent `atomic-like' wavefunctions.  In this system there are actually two ways the overlap can increase, either by decreasing the distance between adjacent wires or by extending the wavefunctions themselves by decreasing the modulation amplitude.  In the long period and/or large amplitude limit, the localized states decouple.  This is akin to the atomic limit and explains the flattening of the dispersion along the $\hat{p}_y$ direction.  In addition, it is important to emphasize that in calculating the transfer integrals, we have made the approximation that the soliton wavefunctions are $k$- independent.  Hence, there is in fact some momentum dependence to the transfer integrals that has been ignored that may become relevant at higher energies and/or short modulation periods.

\section{2D Superlattices}

\textit{Band Structures of 2D BLG Superlattices} --- 
In this section, we will consider the band structure of BLG subject to 2D square superlattices 
with period $\lambda$ in both directions. Specifically, we consider the chessboard like superlattice 
potentials,
\begin{eqnarray}
	&& V_1(x,y)=U_1,\ V_2(x,y)=U_2,\nonumber\\
	&& 0\leq x,y \leq w, {\rm and}\ w \leq x,y \leq \lambda, \nonumber\\
	&& V_{1,2}(x+\lambda,y)=V_{1,2}(x,y+\lambda)=V_{1,2}(x,y),
\end{eqnarray}
and set the average of the superlattice potential to be zero in each supercell. For chemical potential 
superlattices, $U_1=U_2=U$, and electric field superlattices, $U_1=-U_2=U$. 

\begin{figure}[t]
	\centering
	\includegraphics[width=.23\textwidth]{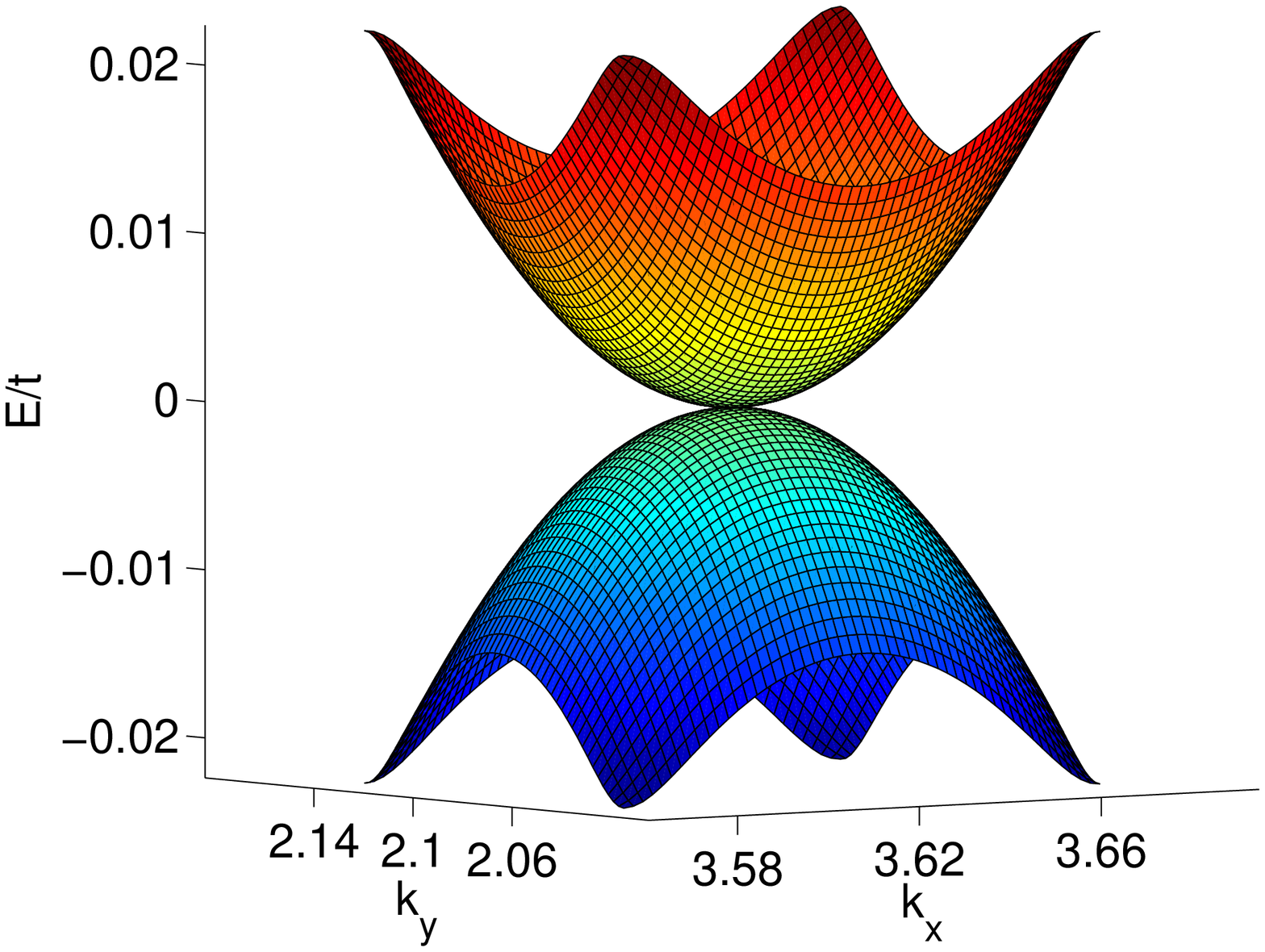}
	\includegraphics[width=.23\textwidth]{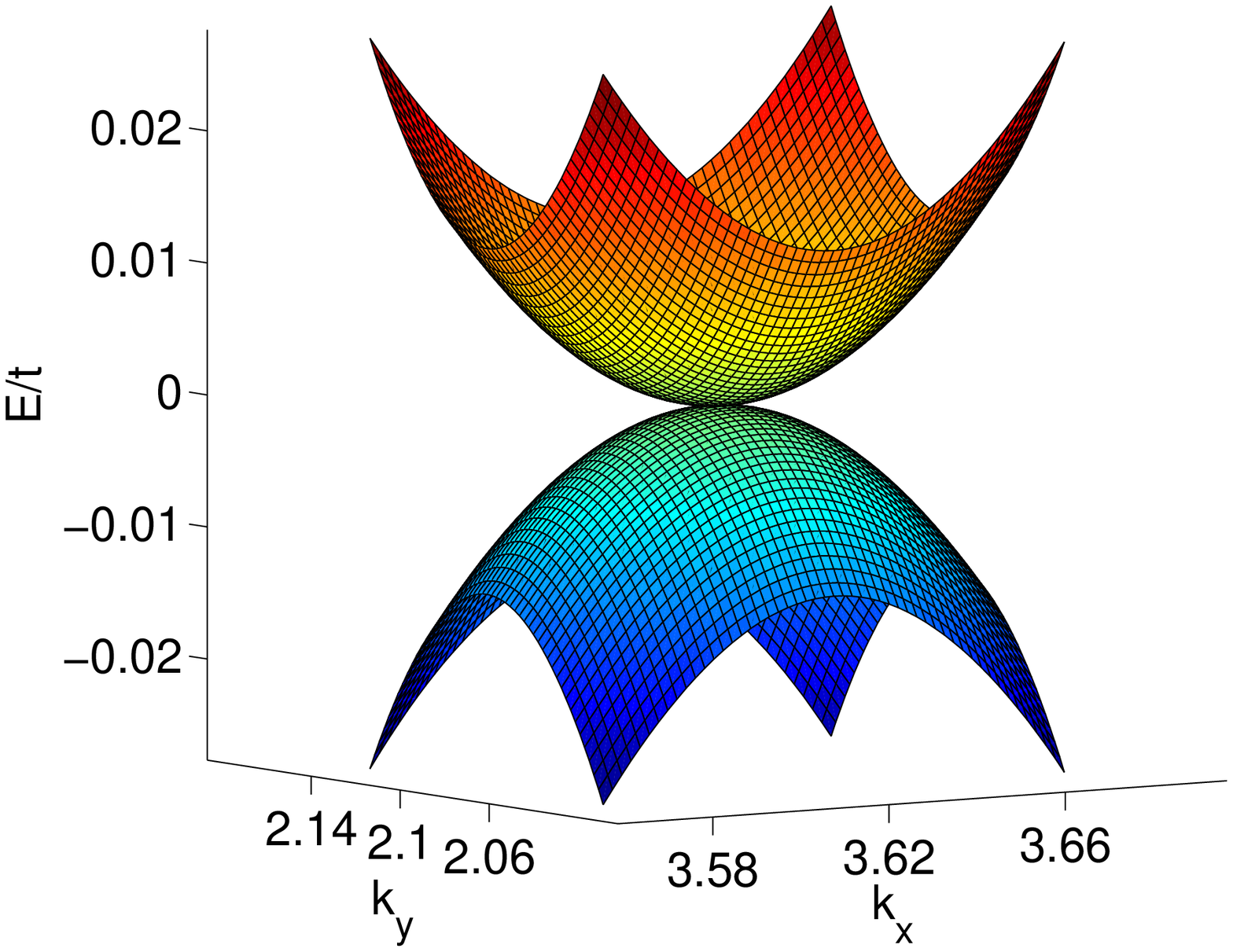}
	\caption{(Color online) Energy spectra for symmetric 2D chemical potential (left) and electric field (right) 
	superlattices. Here, $\lambda=60d$, $w=30d$, $U=0.025t$.}
	\label{fig6}
\end{figure}

Fig. \ref{fig6} shows the highest valence band and the lowest conduction band of 
BLG subject to the above superlattices. Generally, the energy spectra are fully gapped on 
the MBZ boundaries. Around $K$ point, the energy spectra depend on the details 
of the superlattices. When the following symmetry is present, 
\begin{equation}
	\label{Symm}
V_{1,2}(x+\lambda/2,y)=V_{1,2}(x,y+\lambda/2)=-V_{1,2}(x,y), 
\end{equation}
there is no gap opening at the $K$ point. This can be understood from perturbation theory presented in next subsection. 
However, when $w$ deviates from $\lambda/2$, a gap will open at the $K$ point for electric 
field superlattices. For chemical potential superlattices, no gap opens, but the charge neutrality 
point will shift in energy.

\textit{Analysis of Spectrum at $K$ Point} ---
When superlattice potential $U_{1,2}$ is not very large, or the superlattice period $\lambda$ is 
not very small, we can understand the energy spectrum from perturbation theory. Let us focus on 
the electron states at $K$ in a chemical potential superlattice. Up to second order, the energy 
correction from an electron state with momentum ${\bf G}=(n,m)\times 2\pi/\lambda$, where $n$ and $m$ 
are integers, can be directly read off from the $W$-
matrix,
\begin{equation}
	\Delta E^{(2)}_e({\bf G})=\frac{\left|\left(1+e^{2i\theta_{{\bf G}}}\right)U({\bf G})\right|^2}
	{4(0-\varepsilon_e({\bf G}))},
\end{equation}
where $\theta_{{\bf G}}$ is the angle defined earlier. Similarly, the energy correction from a 
hole state with momentum ${\bf G}^{\prime}=(-m,n)\times 2\pi/\lambda$ is
\begin{equation}
	\Delta E^{(2)}_h({\bf G}^{\prime})=\frac{\left|\left(1-e^{2i\theta_{{\bf G}^{\prime}}}\right)
	U({\bf G}^{\prime})\right|^2}{4(0-\varepsilon_h({\bf G}^{\prime}))}.
\end{equation}
Since $\theta_{{\bf G}^{\prime}}=\theta_{{\bf G}}+\pi/2$, 
$|U({\bf G})|=|U({\bf G}^{\prime})|$, and $\varepsilon_e({\bf G})=-\varepsilon_h({\bf G}^{\prime})$, 
the contribution from the above two states will cancel each other. Most importantly, due to the 
fourfold rotation symmetry, it is always possible to find such a pair of electron-hole states whose 
angles differ by $\pi/2$, which makes the cancellation exact even an infinite number of states are 
taken into account. Therefore, the fourfold rotation symmetry 
dictates the second order energy correction to states at $K$ to be zero, and this second order result 
does {\it not} depend on the details of the superlattice. 
The above argument also applies for electric field superlattices.

For symmetric superlattices, chemical potential or electric field, there is no third order energy 
corrections from the superlattice potential due to the fact that $U({\bf G})=0$ if either $n$ or $m$ 
is an even integer. Breaking of the symmetry (\ref{Symm}), or more generally the symmetry
associated with particle-hole transformation followed by translation by $\lambda/2$, 
will lead to the modification of the energy 
spectrum. Numerically, we have observed, in chemical potential superlattices, the charge neutrality point 
will slightly shift in energy, a result we appear to recover in third order perturbation theory. In
electric field superlattices, breaking of ${\cal P}$ is numerically found to lead to a gap at the ${\bf K}$ point.
Further details of the band structure of 2D superlattices will be presented in a future publication.